\documentclass[12pt]{article}
\usepackage{amssymb}
\usepackage{amsmath}
\usepackage{apacite}
\usepackage{latexsym}
\usepackage{epsfig}
\usepackage{graphicx}
\usepackage{subfigure}
\usepackage{wasysym}
\usepackage{threeparttable}
\usepackage{natbib}
\usepackage{color}
\usepackage{epstopdf}
\usepackage{bm}
\usepackage{float}
\usepackage{todonotes}
\usepackage{verbatim}
\usepackage{booktabs}
\usepackage{tabularx}
\usepackage{makecell}
\usepackage{longtable}
\usepackage{array}

\usepackage{hyperref}

\def\dfrac#1#2{{\displaystyle{#1\over#2}}}

\def\boxit#1{\vbox{\hrule\hbox{\vrule\kern6pt
          \vbox{\kern6pt#1\kern6pt}\kern6pt\vrule}\hrule}}

\def\bse{\begin{eqnarray*}}
\def\ese{\end{eqnarray*}}
\def\be{\begin{eqnarray}}
\def\ee{\end{eqnarray}}
\def\bq{\begin{equation}}
\def\eq{\end{equation}}
\def\bse{\begin{eqnarray*}}
\def\ese{\end{eqnarray*}}

\begin{document}

\thispagestyle{empty}
\baselineskip=28pt

\begin{center}
{\LARGE{\bf Topic Modeling for Free-Response Text Data from a Complex Survey}}

\end{center}

\baselineskip=12pt

\vskip 2mm
\begin{center}
Namitha V. Pais\footnote{(\baselineskip=10pt to whom correspondence should be addressed)
Cropping Systems and Water Quality Research Unit,
USDA-ARS,
Columbia, MO USA, npmqx@missouri.edu}, Scott H. Holan \footnote{\baselineskip=10pt Department of Statistics, University of Missouri, Columbia, MO, USA.
Research and Methodology Directorate, U.S. Census Bureau, Washington D.C., USA, holans@missouri.edu}
and Paul A. Parker\footnote{\baselineskip=10pt Department of Statistics, University of California Santa Cruz, CA, USA, paulparker@ucsc.edu}
\\
\end{center}
%
%
%
%
\vskip 4mm
\baselineskip=12pt
\begin{center}
{\bf Abstract}
\end{center}

Topic Modeling is a popular statistical tool commonly used on textual data to identify the hidden thematic structure in a document collection based on the distribution of words. Additionally, it can be used to cluster the documents, with clusters representing distinct topics. The Mixture of Unigrams (MoU) is a standard topic model for clustering document-term data and can be particularly useful for analyzing open-ended survey responses to extract meaningful information from the underlying topics. However, with complex survey designs, where data is often collected on individual (document) characteristics, it is essential to account for the sample design in order to avoid biased estimates. To address this issue, we propose the MoU model under informative sampling  using a pseudolikelihood to account for the sample design in the model by incorporating survey weights. We evaluate the effectiveness of this approach through a simulation study and illustrate its application using two datasets from the American National Election Studies (ANES). We compare our pseudolikelihood-based MoU model to the traditional MoU and assess its effectiveness in extracting meaningful topics from survey data. Additionally, we introduce a hierarchical Mixture of Unigrams (hMoU) accounting for informative sampling where topic proportions are defined as functions of document-level fixed and random effects. We demonstrate the effectiveness of the proposed model through an application to ANES data comparing topic proportions across respondent-level factors such as gender, race, age group, and state.

\baselineskip=12pt
\par\vfill\noindent
\par\medskip\noindent
\clearpage\pagebreak\newpage \pagenumbering{arabic}
\baselineskip=24pt

\section{Introduction}

National data collection agencies conduct numerous surveys that often include open-ended, free-response text data. For example, the Federal Register Notices (FRNs) receive submissions in the form of text comments, where users provide feedback on various regulations and important issues. Health agencies, such as the Food and Drug Administration (FDA), also leverage data from Electronic Health Records (EHR) and medical claims data to make regulatory decisions. The American National Election Studies (ANES) include responses to open-ended political knowledge questions \citep{debell2013harder}, while social economics surveys from the National Bureau of Economic Research (NBER) contains responses on topics like income and estate taxation \citep{stantcheva2022eliciting}. These information-rich unstructured texts capture a variety of responses that cannot be easily captured in one or more closed-ended questions.

Traditionally, these open-ended responses are analyzed using labour-intensive human coding \citep{lupia2018improve}. A more recent automated alternative  the use of topic models \citep{roberts2014structural,pietsch2018topic}, large language models (LLMs) \cite{jansen2023employing,cibelli2024semi} and neural networks \citep{parker2023computationally} for the analysis of such open-ended responses.
Topic Models in particular, are a popular choice to model textual data as often, the sets of words observed in texts represent a coherent theme or topic. Topic models analyze the words in the texts (documents) to discover latent topic structure, and do not require any prior information on the labeling of the documents. Many topic modeling algorithms have been developed over time including non-negative matrix factorization \citep{yan2013learning}, Mixture of Unigrams (MoU) \citep{nigam2000text}, Latent Dirichlet Allocation (LDA) \citep{blei2003latent}, and Structural Topic Models \citep{roberts2013structural}. Among the topic models MoUs are one of the simplest and most efficient tools for clustering textual data, as they assume that documents related to the same topic have similar distributions of words. Therefore, MoU can be used to analyze the free response text data to uncover the hidden topics in the texts and cluster the responses based on these topics. While the implementation of topic models on the open ended text data from surveys can provide valuable insights, it is important to consider consider the sampling design in surveys in order to avoid biased results.

In complex surveys, where individuals are sampled with different probabilities, it is important to account for the sampling design in the model to avoid biased estimates.  For example, in data where there exists a relationship between a unit's probability of selection and the response of interest, often knows as \textit{informative sampling}, failure to account for the sampling design can lead to substantial bias. There are several ways to account for the sampling design in  the model. \cite{parker2023unit} discusses modern methods that handle informative sampling for unit-level models. One way is to parameterize the the sampling design into the model using inclusion indicator variables \citep{pfeffermann1998parametric}. Another way is to use the survey weights to re-weight the  likelihood contribution from each unit known as the pseudo-likelihood approach  \citep{binder1983variances,skinner1989domain,savitsky2016bayesian}.

In the MoU model under informative sampling, the sample design can be incorporated in the model by using the survey weights. In this paper, we discuss the MoU model accounting for informative sampling by introducing  the pseudo-likelihood at the data stage of the model. We also introduce a hierarchical MoU model accounting for informative sampling, where the topic proportions  are modeled using document level fixed and random effects. We illustrate the application of these methods using the ANES post-election response data, chosen in part for its public availability. However, our method could readily be applied, with little to no modification, to confidential micro-data coming from complex surveys.

The remainder of this paper proceeds as follows. Section~\ref{method} provides details of the MoU model accounting for informative sampling. Section~\ref{simstudy} discusses a simulation study to assess the effectiveness of our method. Section~\ref{hMoU_IS} introduces the hierarchical Mixture of Unigrams (hMoU) model. In Section~\ref{data_anes},  we illustrate the application of the MoU and hMoU model using two datasets from the ANES study. Finally, in Section~\ref{disc}, we summarize our method and provide concluding remarks.

\section{Methodology}\label{method}
\subsection{Background and Notation}\label{MoU}
We begin by introducing the notation to describe textual data. Suppose we have a corpus $\mathcal{C}$ consisting of $M$ documents represented as  $\mathcal{C} =\{ \boldsymbol{w}_1, \ldots, \boldsymbol{w}_M $\}. Each of the $d^{th}$ document is represented as a collection of $N_d$ words, i.e.,
$\boldsymbol{w}_d=\{\boldsymbol{w}_{d,1},\ldots,\boldsymbol{w}_{d,N_d}\}$  for $d=1,\ldots,M$. The vocabulary from which words are drawn is indexed by $\{1,2,\ldots,V\}$ such that each word $\boldsymbol{w}_{d,n}$ is represented by a $V$-dimensional unit (basis) vector, i.e., $\boldsymbol{w}_{d,n} =(w_{d,n}^1,\ldots,w_{d,n}^V)'$ where,
\begin{align*}
w_d^u &= 
\begin{cases}
1 & \text{if } u = v, \\
0 & u \neq v,
\end{cases}
\label{wordseq}
\end{align*}
when $\boldsymbol{w}_{d,n}$ corresponds to  the $v^{th}$ word from the vocabulary, for $n=1,\ldots,N_d$ and $d=1,\ldots,M$. The MoU model under the non-informative survey design assumes that each of the $M$ documents is generated by first choosing a topic $z_d$ from a multinomial distribution and then generating $N_d$ words independently from the conditional multinomial distribution for $d=1,\ldots,M$. The generative process assumed by the MoU model is given by
\begin{eqnarray*}
\boldsymbol{w}_{d,n} \mid \boldsymbol{z}_d, \boldsymbol{\phi} & \sim & Mult(1, \boldsymbol{\phi}^{z_d}) \quad \text{for} \quad n=1,\ldots,N_d \quad \text{and} \quad d=1,\ldots,M. \\
\boldsymbol{z}_d \mid \boldsymbol{\theta} & \sim & Mult(1,\boldsymbol{\theta}) \quad \text{for} \quad d=1,\ldots,M. \\
\boldsymbol{\theta} \mid \boldsymbol{\alpha} & \sim & Dir(\boldsymbol{\alpha}) \\
\boldsymbol{\phi}^{j} & \sim & Dir(\boldsymbol{\eta}) \quad \text{for} \quad j=1,\ldots,J.
\end{eqnarray*}
Here,  the number of topics $J$ is assumed to be fixed and known. The parameters associated with the generative process are the vector of topic proportions $\boldsymbol{\theta}$ assumed to come from symmetric Dirichlet distribution with concentration parameter $\alpha$ and the matrix of word probabilities $\boldsymbol{\phi}$ where for each topic $j$ the  vector of word probabilities $\boldsymbol{\phi}^j$ is assumed to come from a symmetric Dirichlet distribution with concentration parameter $\eta$ for $j=1,\ldots,J$. Under this mixture model framework, the probability of the $d^{th}$ document is given by
 \begin{equation*}
     p(\boldsymbol{w}_d)=\sum_{z} p(\boldsymbol{z} \mid \boldsymbol{\theta}) \prod_{n=1}^{N_d} p(\boldsymbol{w}_{d,n} \mid \boldsymbol{z},\boldsymbol{\phi}),
 \end{equation*}
for $d=1,\ldots,M$. The estimated model can be used to identify the underlying topics and their estimated proportions, obtain the distribution of words across each topic, and cluster the documents into $J$ groups based on the topic assignments.

\subsection{Proposed Model}\label{MoUIS}
We now introduce the MoU model accounting for informative sampling. While fitting models with unit-level survey data, we work with samples collected using complex sampling designs with known survey weights where, there may exist a dependence between the probabilities of selection, and the underlying thematic structure of the text data. Hence, it is important to account for this relationship in order to identify the underlying topics correctly and avoid biased estimates. One  way to account for informative sampling is to use the pseudo-likelihood where the reported survey weights are used to exponentially re-weight the data likelihood contribution of each document \citep{parker2023unit}. Under the MoU model accounting for informative sampling, the pseudo-likelihood associated with the the $d^{th}$ document is given by
\begin{equation*}
p({\boldsymbol{w}_{d}}\mid \boldsymbol{\theta},\boldsymbol{\phi}) = \Bigg[\sum_{z} p(\boldsymbol{z} \mid \boldsymbol{\theta}) \prod_{n=1}^{N_d} p(\boldsymbol{w}_{d,n} \mid \boldsymbol{z},\boldsymbol{\phi})\Bigg]^{\omega_d},
\end{equation*}
for $d=1,\ldots,M$. Thus, under the pseudolikelihood-based MoU model accounting for informative sampling, the model can be written as

\begin{eqnarray*}
p({\boldsymbol{w}_{d}}\mid \boldsymbol{\theta},\boldsymbol{\phi}) & = & \Bigg[\sum_{\boldsymbol{z}} p(\boldsymbol{z} \mid \boldsymbol{\theta}) \prod_{n=1}^{N_d} p(\boldsymbol{w}_{d,n} \mid \boldsymbol{z},\boldsymbol{\phi})\Bigg]^{\omega_d} \quad d=1,\ldots,M. \\
\boldsymbol{z} \mid \boldsymbol{\theta} & \sim & Mult(1,\boldsymbol{\theta}) \\
\boldsymbol{\theta} \mid \boldsymbol{\alpha} & \sim & Dir(\boldsymbol{\alpha}) \\
\boldsymbol{\phi}^{j} & \sim & Dir(\boldsymbol{\eta})  \quad j=1,\ldots,J.
\end{eqnarray*}
Within the pseudo-likelihood, we use the scaled survey weights, $\omega_d$, such that the weights sum to the sample size.

\section{Simulation Study}\label{simstudy}
To evaluate the performance of our model, we conduct a simulation study. We generate a population consisting of $M=10,000$ documents, where each document is represented by a collection of words with length $N_d$ assumed to come from a Poisson distribution with mean $ \lambda=30$. The vocabulary size is set to $V=6$ and the underlying latent topic structure is set to $J=3$ topics. To obtain an informative sample, we consider a subsample of $M_{sub}=100$ documents (constituting  $1\%$ of the total population) from the population.  The documents are sampled such that the documents belonging to one of the $J$ topics are over-sampled, while the remaining documents are under-sampled. This approach results in an informative sample from the population.  We then fit our model using MCMC with 3,000 iterations, discarding the first 500 iterations as burn-in. All the models are fit using the Hamiltonian Monte Carlo (HMC) algorithm \citep{neal2012mcmc} in Stan using the \textit{cmdstanr} package in R \citep{cmdstanr}. We repeat the sampling and estimation procedure for $K=100$ times. Convergence is assessed via trace plots of the sample Markov chains, with no lack of convergence detected. To compare the model performance, we also fit the standard MoU without accounting for the informative survey design. We consider three forms of assessment to compare the two models. First, we consider the root mean squared error (RMSE) of our point estimates given by 
\begin{equation*}
    RMSE= \sqrt{\sum_{k=1}^K \dfrac{(\hat \theta_k -\theta)^2}{K}},
\end{equation*}
where $\theta$ represents the true population quantity of interest and $\hat \theta_k$  represents an estimate for sample dataset k, for $k =1,\ldots,K$. Similarly, we also consider absolute bias given by 
\begin{equation*}
    Bias= \left|\dfrac{1}{K} \sum_{k=1}^K \hat \theta_k - \theta \right|.
\end{equation*}
We also consider the interval score \citep{gneiting2007strictly} for the $95\%$ credible interval estimates
\begin{equation*}
    IS= \dfrac{1}{K} \sum_{k=1}^K \Bigg\{ (u_k-l_k)+\dfrac{2}{\alpha} (l_k-\theta) I(\theta<l_k)+\dfrac{2}{\alpha} (\theta-u_k) I(\theta>u_k)\Bigg\},
\end{equation*}
where $\alpha=0.05$, $u_k$ is the upper bound of the interval and $l_k$ is the lower bound of the interval for sample dataset $k$ for $k=1,\ldots,K$. For the interval score, a lower score is desirable.  Table~\ref{tab:combined_table} compares the two modes based on the three evaluation metrics. For the parameter matrix $\boldsymbol{\phi}$, we observe that both models have similar RMSE, absolute bias, and interval score values, with slightly lower absolute bias and interval score when accounting for informative sampling. This suggests that the estimation of $\boldsymbol{\phi}$, representing the matrix of word probabilities across each topic, is not significantly impacted by informative sampling. For the topic proportion vector $\boldsymbol{\theta}$, there is a substantial reduction in RMSE, absolute bias, and interval score when accounting for informative sampling. The standard MoU model tends to assign a higher topic proportion to the oversampled topic while underestimating the proportions of other topics, resulting in inaccurate topic representations. Therefore, accounting for the sampling design is essential to obtain more reliable and accurate estimation of topic proportions.

\begin{table}
\begin{tabular}{lcccccc}
\toprule
\textbf{Parameter} & \multicolumn{3}{c}{\makecell{\textbf{MoU under the}\\\textbf{non-informative survey design}}} & \multicolumn{3}{c}{\makecell{\textbf{MoU accounting for}\\\textbf{informative sampling}}} \\
\cmidrule(lr){2-4} \cmidrule(lr){5-7}
\cmidrule(lr){2-4} \cmidrule(lr){5-7}
 & \textbf{RMSE} & \textbf{Abs Bias} & \textbf{Interval Score} & \textbf{RMSE} & \textbf{Abs Bias} & \textbf{Interval Score} \\
\midrule
$\boldsymbol{\phi}$ & 0.0126 & 0.0023  & 0.0013 & 0.0127 & 0.0021 & 0.0016 \\
$\boldsymbol{\theta}$ & 0.2419 & 0.2202 & 0.1421 & 0.0458 & 0.004 &  0.0062 \\
\bottomrule 
\end{tabular}
\label{tab:combined_table}
\caption{Evaluation metrics comparing the two models on the simulated data.}
\end{table}

\begin{figure}
\begin{center}
\includegraphics[scale=0.6]{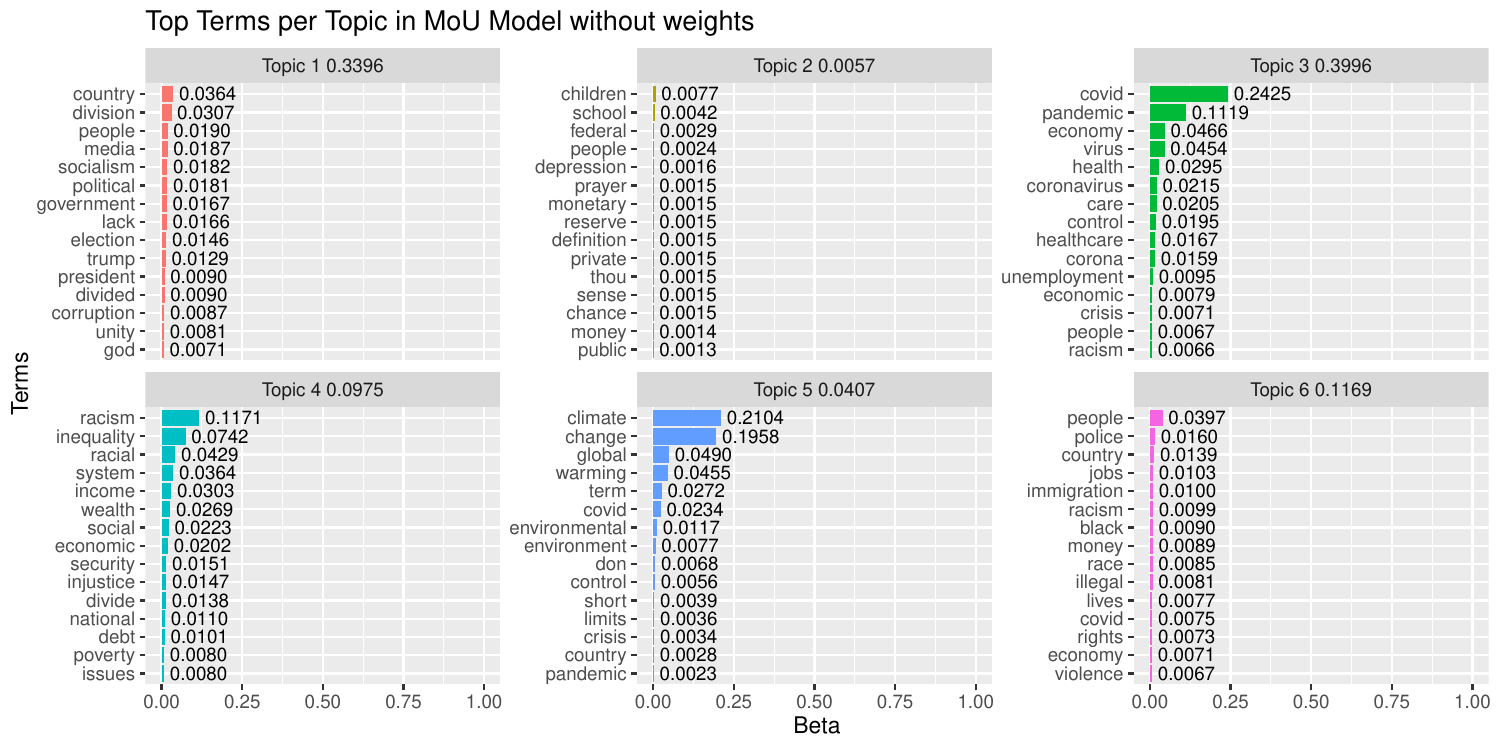}
\caption{Estimated distribution of the top 15 words over each of the $J=6$ latent topics obtained from the MoU model under the non-informative survey design, for the ANES 2020 data.}
\label{fig:beta_plot_2020_1}
\end{center}
\end{figure}

\begin{figure}
\begin{center}
\includegraphics[scale=0.6]{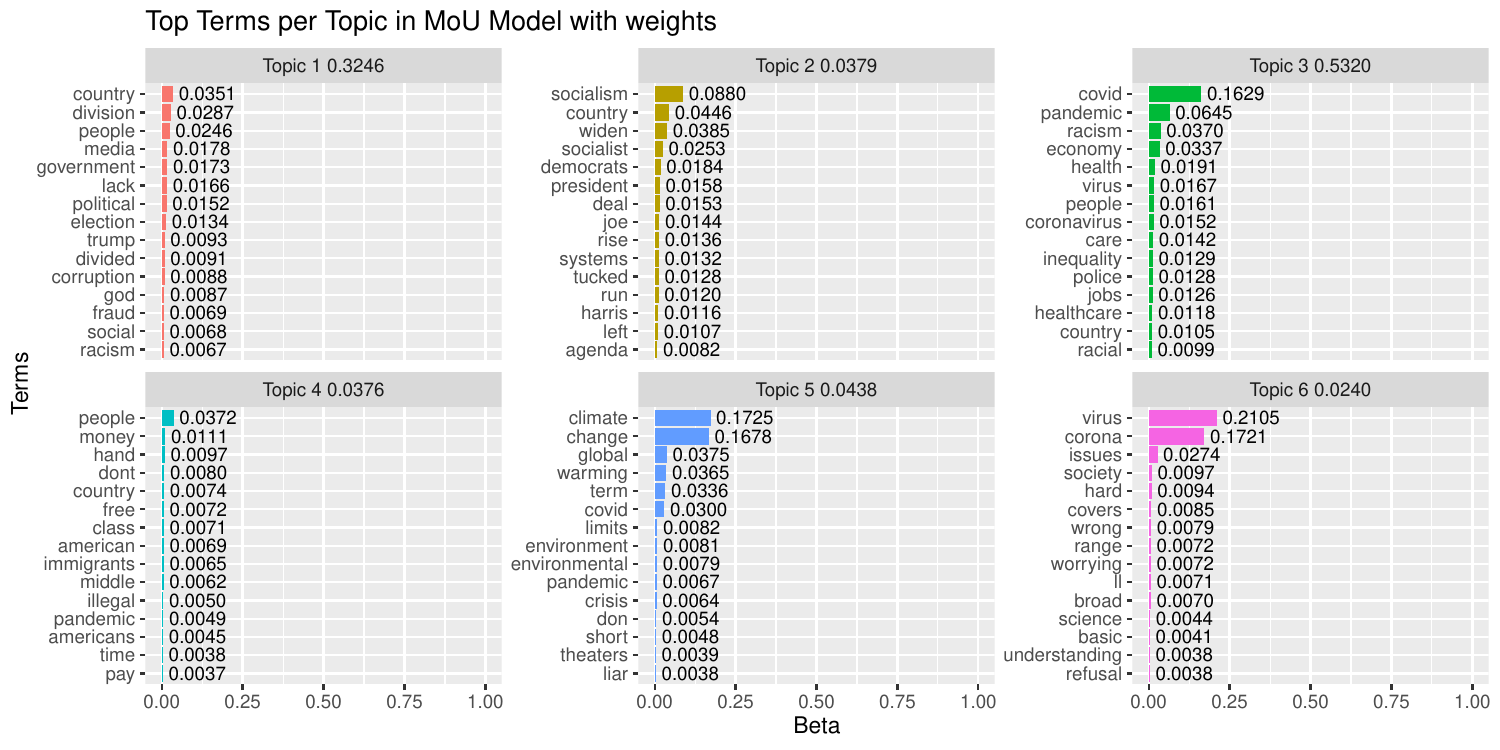}
\caption{Estimated distribution of the top 15 words over each of the $J=6$ latent topics obtained from the MoU model accounting for informative sampling, for the ANES 2020 data.}
\label{fig:beta_plot_2020_2}
\end{center}
\end{figure}

\section{Hierarchical Mixture of Unigrams  accounting for informative sampling}\label{hMoU_IS}
In addition to estimating the overall topic structure, it may be important to examine how topic structures vary across different respondent (or document) groups. For instance, in post-election responses, it is useful to see how topic proportions differ based on respondent-level factors such as gender, race, age group, or state. To address this, we propose a hierarchical MoU model that accounts for informative sampling where the topic proportions are defined as a function of document level fixed and random effects. The hierarchical MoU model accounting for informative sampling is given by
\begin{eqnarray*}
p({\boldsymbol{w}_{d}}\mid \boldsymbol{\theta_d},\boldsymbol{\phi}) &=& \Bigg[\sum_{z_d} p(\boldsymbol{z_d} \mid \boldsymbol{\theta_d}) \prod_{n=1}^{N_d} p(\boldsymbol{w}_{d,n} \mid \boldsymbol{z_d},\boldsymbol{\phi})\Bigg]^{\omega_d}, \quad d=1,\ldots,M\\
\boldsymbol{z_d} \mid \boldsymbol{\theta_d} &\sim& Mult(1,\boldsymbol{\theta_d}), \quad d=1,\ldots,M \\
\theta_{d,j} &=& \frac{\exp(\xi_{d,j})}{1+\sum_{l=1}^{(J-1)} \exp(\xi_{d,l})}, \quad j=1,\ldots,J-1\\
\xi_{d,j} &=& \boldsymbol{X}_d^T\boldsymbol{\beta}_j + \boldsymbol{\psi}_d^T\boldsymbol{\gamma}_j, \quad d=1,\ldots,M\\
\boldsymbol{\gamma}_j \mid \sigma^2_{\gamma} &\sim& N_r(\boldsymbol{0}_r,\sigma^2_{\gamma_j} \boldsymbol{I}_r), \quad j=1,\ldots,J-1\\
\boldsymbol{\beta}_j &\sim& N_p(\boldsymbol{0}_p, \sigma^2_{\beta} \boldsymbol{I}_p), \quad j=1,\ldots,J-1\\
\sigma^2_{\gamma} &\sim& IG(a,b) \\
\boldsymbol{\phi}^{j} &\sim& Dir(\boldsymbol{\eta}), \quad j=1,\ldots,J.
\end{eqnarray*}
Here, the vector $\boldsymbol{X}_d$ represents a \( p \)-dimensional set of covariates associated with the $d^{th}$ document, and $\boldsymbol{\beta}_j$ is a \( p\)-dimensional vector of fixed effects corresponding to the $j^{the}$ topic. The vector $\boldsymbol{\psi}_d$ corresponds to the \( r \)-dimensional incidence vector associated with the $d^{th}$ document, and $\boldsymbol{\gamma}_j$ is an \( r \)-dimensional vector of random effects associated with the $j^{th}$ topic. It should be noted that we constrain $\boldsymbol{\beta}_J$ and $\boldsymbol{\eta}_J$ to be equal to the zero vector for identifiability. We put a  \( p \)-dimensional Normal prior distribution on the vector of fixed effects $\boldsymbol{\beta}_j$ with mean vector $\boldsymbol{0}$ and variance-covariance matrix  $\sigma^2_\beta \boldsymbol{I}$, where $\sigma^2_\beta$ is assumed to be a hyperparamter. Similarly, we put a  \(r \)-dimensional Normal prior on the vector of random effects $\boldsymbol{\gamma}_j$ with mean vector $\boldsymbol{0}$ and variance-covariance matrix $\sigma^2_\gamma \boldsymbol{I}$, where $\sigma^2_\gamma$ is assumed to come from gamma distribution with shape and rate parameters as $a$ and $b$, treated as hyperparamters. The hyperparameters are set to yield a vague prior distribution with $a=b=0.1$ and $\sigma^2_\beta=1000$. This framework provides distinct topic proportions for each combination of respondent-level fixed and random factors, allowing for comparisons across different document level groups. In the next section, we present and discuss the results from the models outlined in Sections~\ref{MoU}, \ref{MoUIS}, and \ref{hMoU_IS}, applied to the ANES data.

\section{ANES Illustration}\label{data_anes}
We apply the methods discussed in Sections~\ref{MoU},~\ref{MoUIS}, and~\ref{hMoU_IS}  to two datasets from the American National Election Studies (ANES) survey. We use the time series study data which records various responses during both pre- and post-election. We only consider the post-election data, which contains a variety of free response questions. In particular, we analyze the free responses from the question ``Which among mentions is the most important  problem the country faces?". We consider the free response texts for this question for the years 2016 and 2020 that are responded to in English. Tables \ref{tab:anes_2016_responses}
and \ref{tab:anes_2020_responses} present snapshots of responses from 2016 and 2020, respectively. The initial data pre-processing includes removing standard list of stop words, characters and numbers along with some colloquial spelling corrections. After the data pre-processing, to determine the suitable number of topics $J$ on each dataset, we run our method (Section~\ref{MoUIS}) for different values of $J$ starting from two until the proportion of the least significant topic is less than  $1\%$. We run both the MoU models discussed in Section~\ref{MoU} and Section~\ref{MoUIS} to assess the impact
of incorporating the informative sampling design into the model.
Next, we present the results from our hierarchical MoU model (Section ~\ref{hMoU_IS}) and compare the estimated topic proportions accross various groups. The topic proportions in the hierarchical MoU model are defined as a function of fixed effects, including gender, race, and age group, along with state-level random effects.

\begin{table}[h!]
\centering
\renewcommand{\arraystretch}{1.5}
\resizebox{\textwidth}{!}{ 
\begin{tabular}{|c|p{15cm}|}
\hline
\textbf{ID} & \textbf{Response} \\
\hline
300008 & violence between community and police. \\
\hline
300034 & division between people. it's awful! \\
\hline
300042 & bringing the country together, working as a country together, not being divisive, caring about each other. \\
\hline
300051 & debt in general. we are so far in debt. everybody else owns us because we are so far in debt. \\
\hline
300053 & keeping jobs in the united states. \\
\hline
300140 & racial divide, people feeling distrust of your fellow american, it's permeating all sectors of our lives. \\
\hline
\end{tabular}}
\caption{Snapshot of open-ended responses from ANES 2016 data.}
\label{tab:anes_2016_responses}
\end{table}

\begin{table}[h!]
\centering
\renewcommand{\arraystretch}{1.5}
\resizebox{\textwidth}{!}{ 
\begin{tabular}{|c|p{15cm}|} 
\hline
\textbf{ID} & \textbf{Response} \\
\hline
200039 & severe political polarization and division that leaves no room for discussion or compromise \\
\hline
200282 & income inequality. wealth distribution \\
\hline
200343 & pollution of air and water supply \\
\hline
200435 & given there is no chance we will transition into any semblance of anarchy (the actual definition, not the distorted view) \\
\hline
200473 & riots of destruction in large cities and lack of respect for others of different color or religion. plus, political manipulation \\
\hline
200480 & liberty to speak without violence and justice for all \\
\hline
\end{tabular}}
\caption{Snapshot of open-ended responses from ANES 2020 data.}
\label{tab:anes_2020_responses}
\end{table}

\begin{figure}
\begin{center}
\includegraphics[scale=0.6]{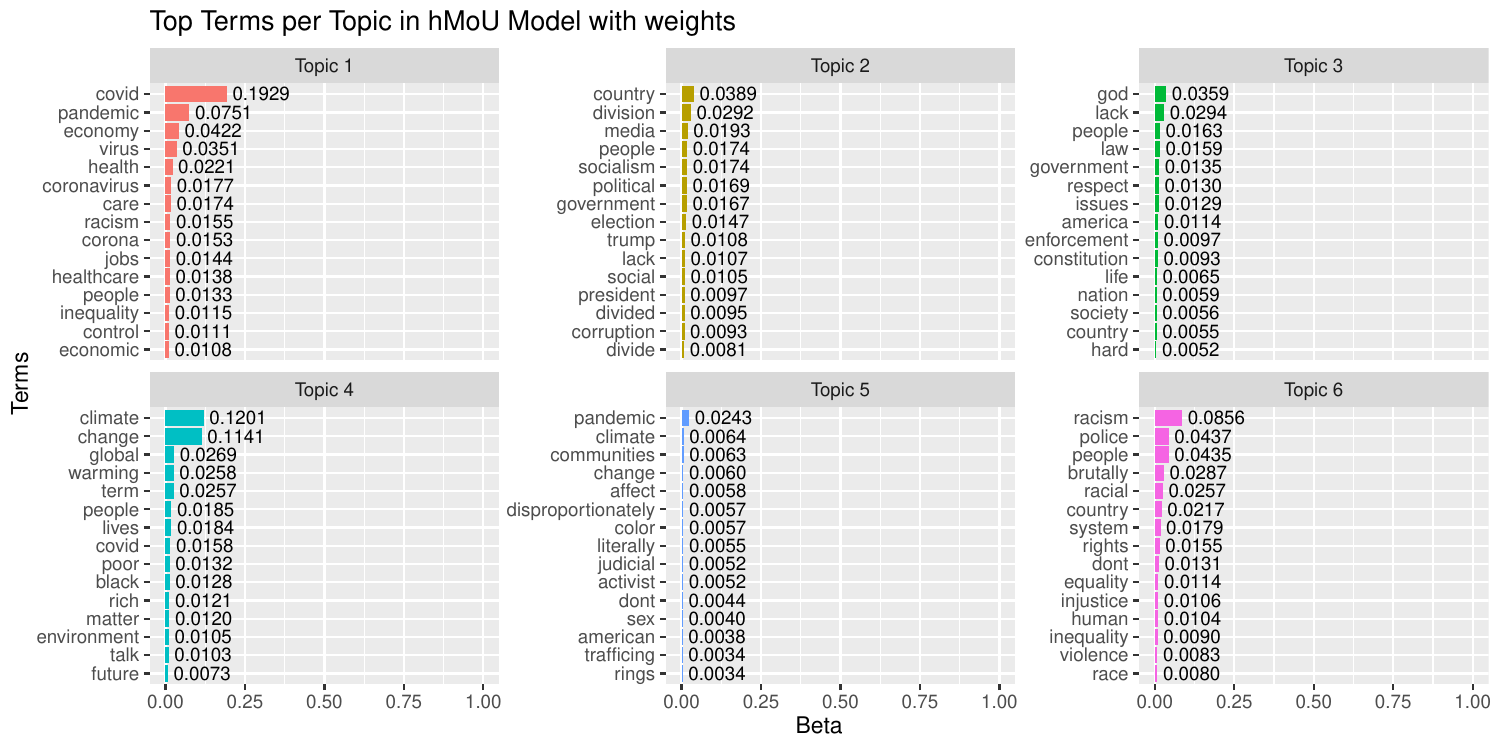}
\caption{Estimated distribution of the top 15 words over each of the $J=6$ latent topics obtained from the hierarchical MoU model under the informative survey design, for the ANES 2020 data.}
\label{fig:hMoU_beta_plot_2020}
\end{center}
\end{figure}
\subsection{Results from the ANES 2020 data}
We analyze the free response texts  from the ANES 2020 data which contains $M= 5389$ responses (or documents) and has vocabulary size  of $V= 2720$. Based on our criterion, the number of topics are chosen to be $J=6$. We implement both the MoU models to evaluate the estimated topic proportions, along with the distribution of the top $15$ words across each topic shown in Figures  \ref{fig:beta_plot_2020_1} and  \ref{fig:beta_plot_2020_2}. From the figures, we can see that the latent topic structure obtained from the free response text depends on the probabilities of selection, as the two models lead to significantly different estimates. For instance, the most frequent topic based on its distribution of words corresponds to the \textit{COVID-19 pandemic, healthcare, and social inequality} as the most important problems in the country. While both the models are able to identify this topic, the estimated topic proportion under the MoU model is $0.3996$ which is substantially different and under estimated in comparison to the estimated topic proportion of $0.5320$ obtained from the MoU model accounting for informative sampling. This is an important finding. Similarly the second most prominent topic which corresponds to \textit{political division and media influence} as the most important problem is over estimated with topic proportion of $0.3396$ in the standard MoU model when compared to the estimated topic proportion of $0.3246$ obtained from the the MoU model accounting for informative sampling.  Thus, it is important to account for this relationship using the MoU model accounting for informative sampling (Figure~\ref{fig:beta_plot_2020_2})  in order to avoid biased estimates  that are obtained when we do not account for informative sampling as seen in Figure~\ref{fig:beta_plot_2020_1}. 

\begin{figure}
\begin{center}
\includegraphics[scale=0.6]{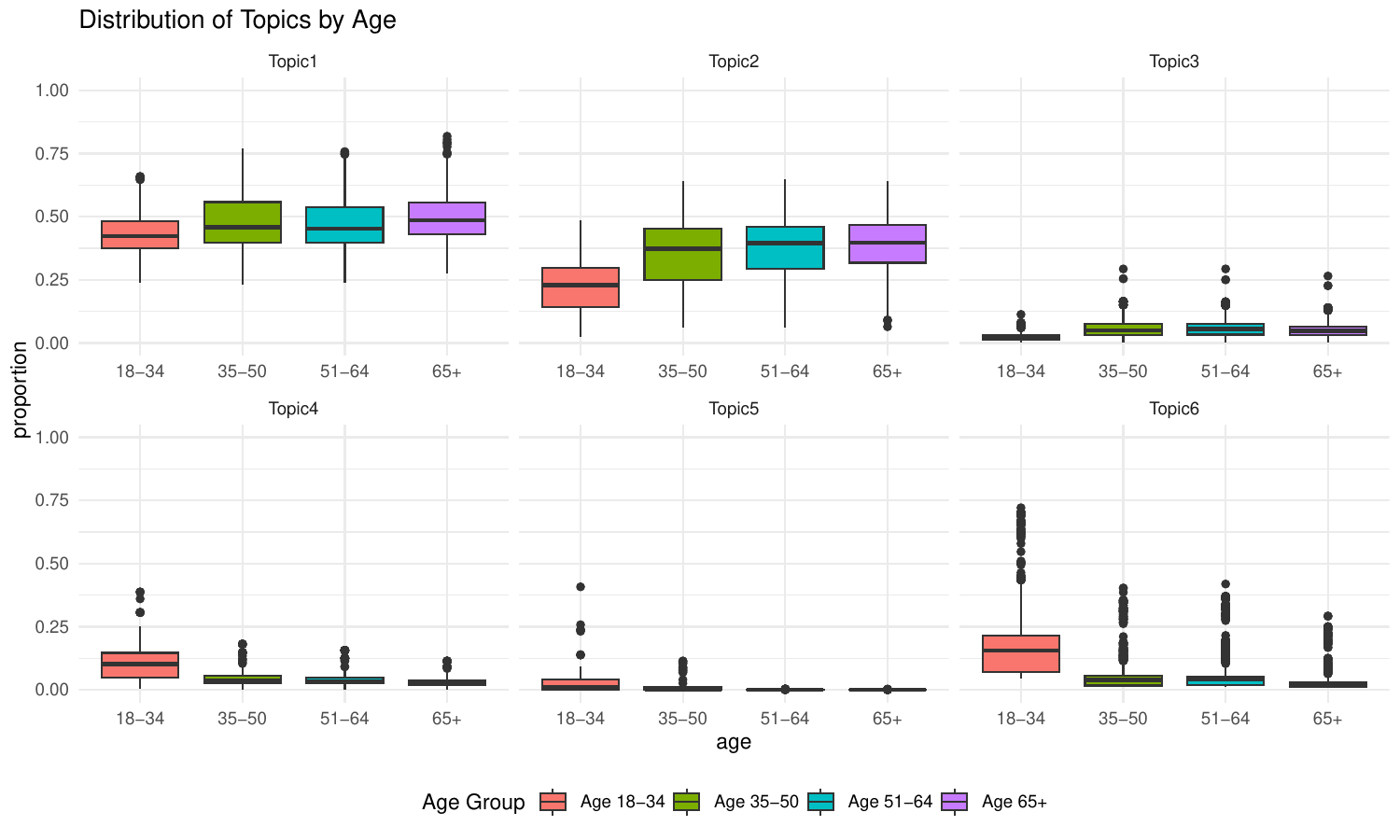}
\caption{Distribution of topic proportions for each of the $J=6$ latent topics obtained from the hMoU model across different age groups for the ANES 2020 data.}
\label{fig:topic_by_age_2020}
\end{center}
\end{figure}

\begin{figure}
\begin{center}
\includegraphics[scale=0.6]{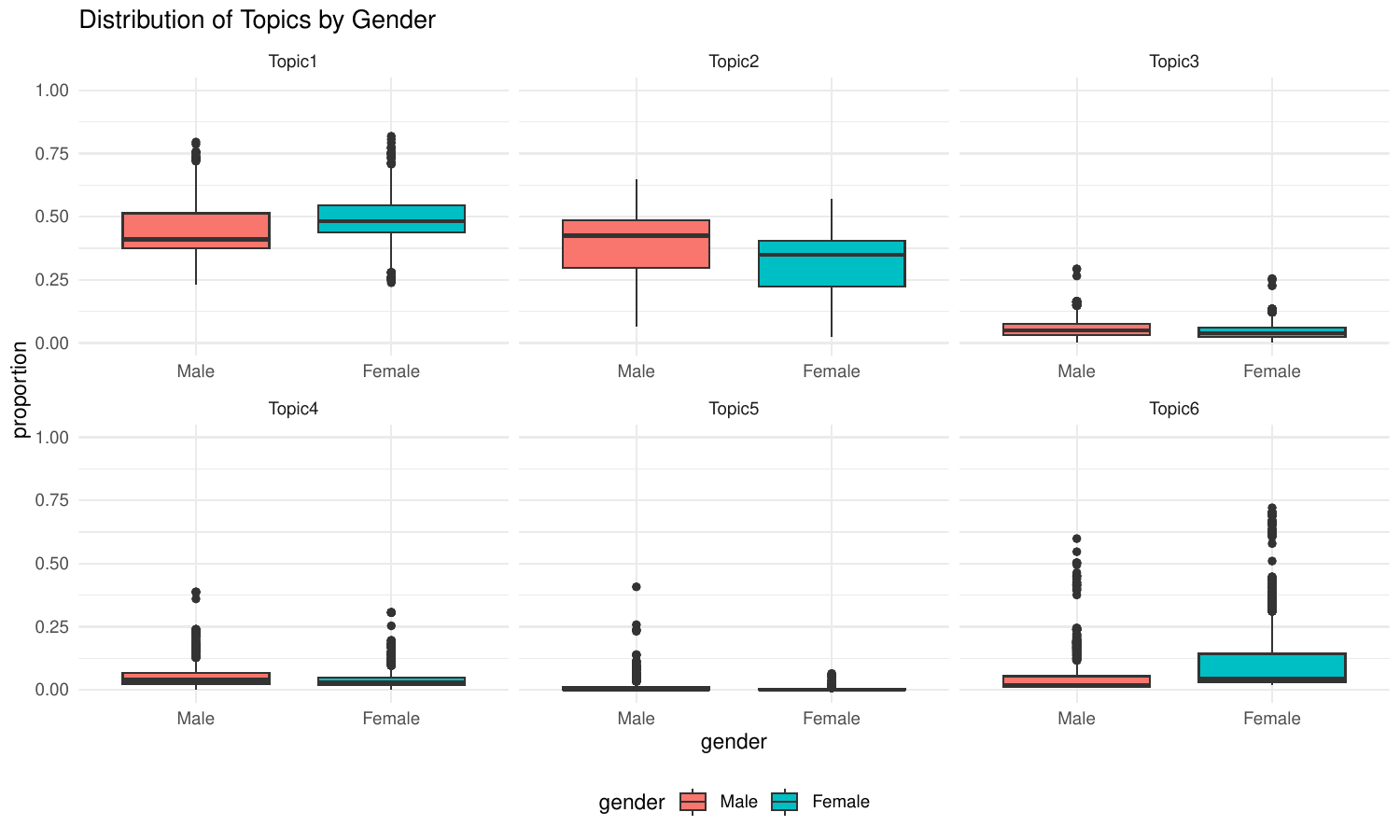}
\caption{Distribution of topic proportions for each of the $J=6$ latent topics obtained from the hMoU model across different genders for the ANES 2020 data.}
\label{fig:topic_by_gender_2020}
\end{center}
\end{figure}

\begin{figure}
\begin{center}
\includegraphics[scale=0.6]{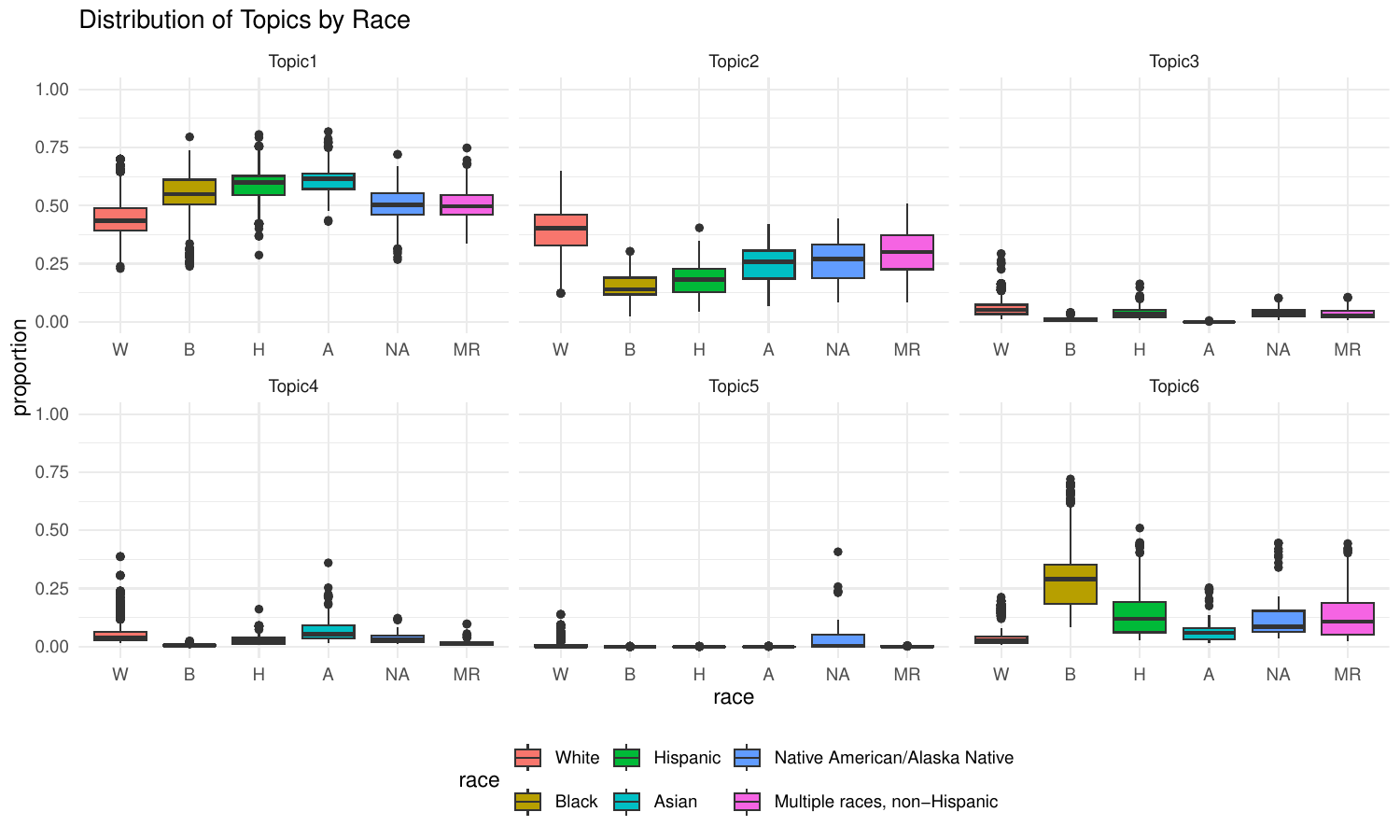}
\caption{Distribution of topic proportions for each of the $J=6$ latent topics obtained from the hMoU model across different race groups for the ANES 2020 data.}
\label{fig:topic_by_race_2020}
\end{center}
\end{figure}

\begin{figure}
\begin{center}
\includegraphics[scale=0.6]{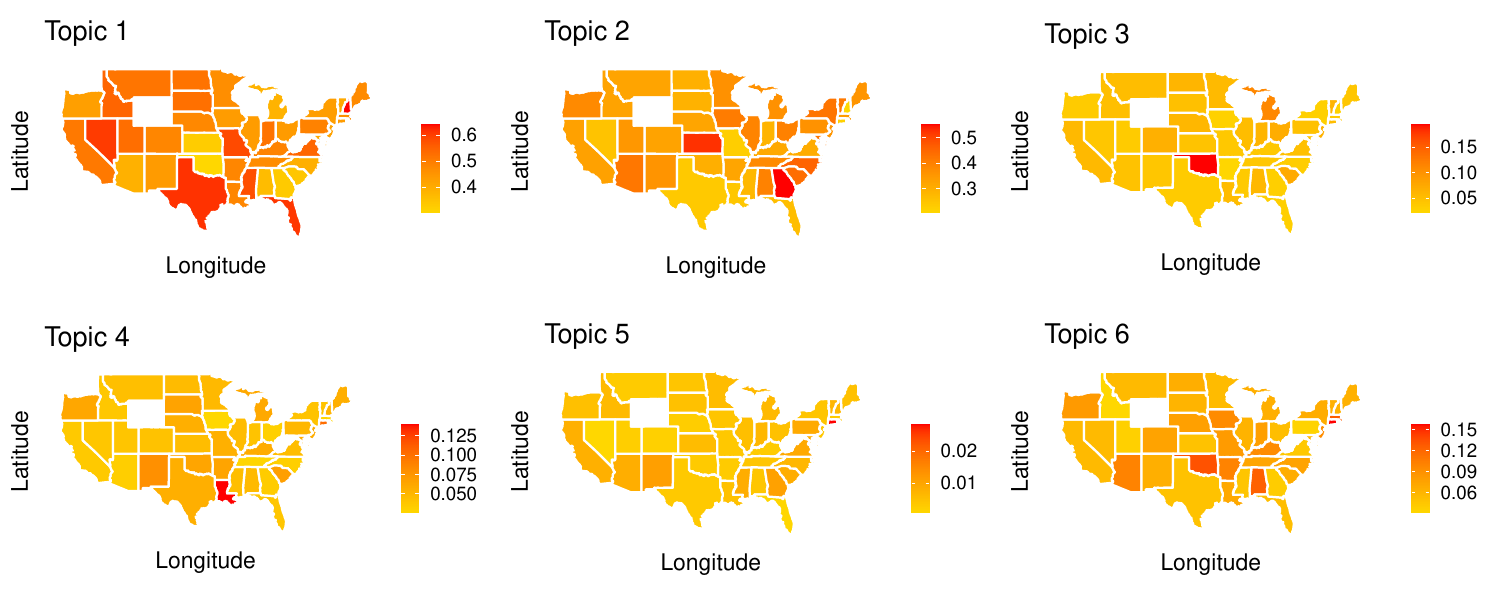}
\caption{Distribution of topic proportions for each of the $J=6$ latent topics obtained from the hMoU model across different states for the ANES 2020 data.}
\label{fig:topic_by_state_2020}
\end{center}
\end{figure}

Furthermore, we also run the hierarchical MoU model, where the the topic proportions are modeled as a function of fixed effects, including respondents age group, gender, and race group, along with state-level random effects. Figure~\ref{fig:hMoU_beta_plot_2020} shows the distribution of the top $15$ words across each topic and 
Figures \ref{fig:topic_by_age_2020}-\ref{fig:topic_by_state_2020} shows the distribution of the $J=6$ topic proportions across various groups including age, gender, race, and state. Similar to the MoU under the informative sampling model (Figure~\ref{fig:beta_plot_2020_2}), the two popular topics correspond to \textit{COVID-19 pandemic, healthcare, and social inequality} and \textit{political division and media influence} (Figure~\ref{fig:hMoU_beta_plot_2020}). Among the age, gender and  race groups, we make the following observations.
Among the age groups, respondents aged 35 and above show higher proportions for the Topics 1 and 2  compared to those aged 18-34. For age group 18-34, Topics 3 and 4 have higher proportions among compared to other age groups. 
Among the gender groups, Topic 1 is more prominent among females than males, while Topic 2 is more prominent among males than females. Across race groups, Topic 2 is most prominent among White respondents.  Topic 6, which is related to racism and injustice, is notably more prominent among the Black race group compared to other race groups. Across states, we observe the following: Topic 1 is most prominent among respondents from Texas and Nevada; Topic 2 is more prominent among respondents from Kansas and Georgia; Topic 3 is most prominent among respondents from Oklahoma; Topic 4 is most prominent among respondents from Louisiana; and Topic 6 is most prominent among respondents from Arkansas and Alabama compared to respondents from other states.

\begin{figure}
\begin{center}
\includegraphics[scale=0.6]{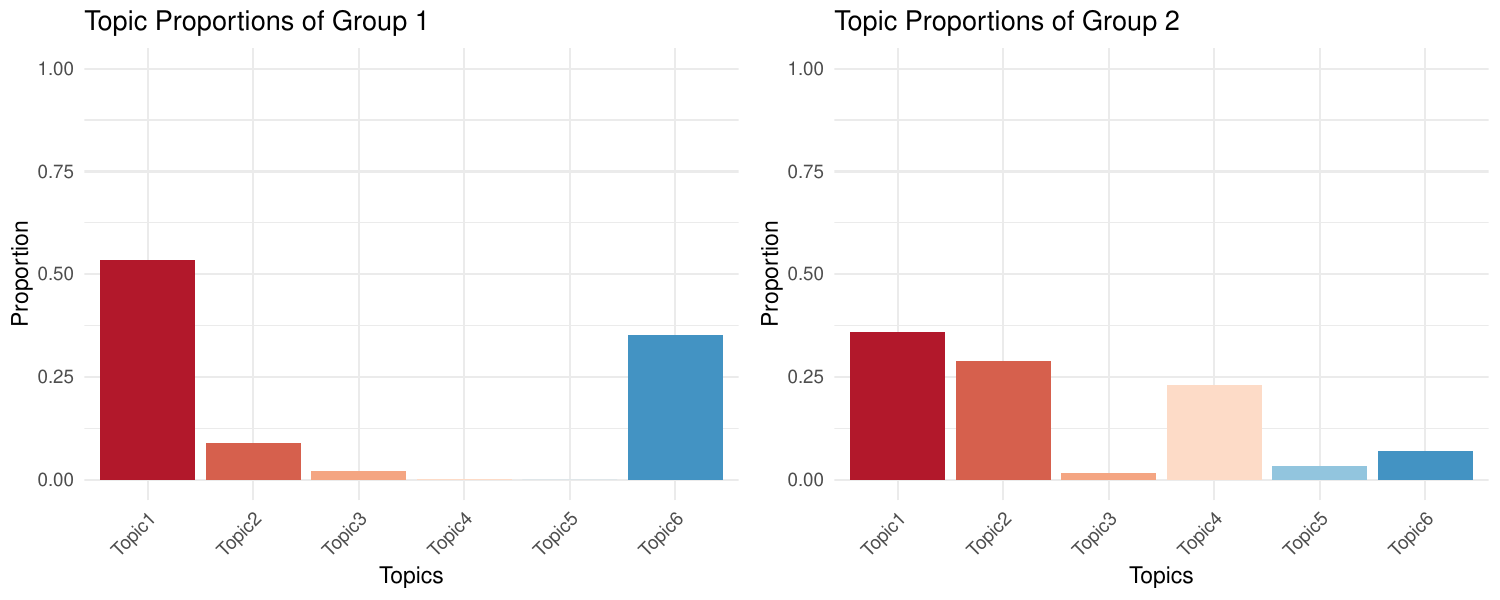}
\caption{Barplots comparing topic proportions of two distinct respondent groups. Group~1 consists of respondents who are Black females aged 35-50 from the state of Georgia and Group~2 consists of respondents who are White males aged 18-34 from the state of California.}
\label{fig:compare_2020}
\end{center}
\end{figure}

In addition to comparing the distribution of topic proportions across respondent-level factors, one can also compare the topic proportions between two distinct groups of respondents. For instance, Figure~\ref{fig:compare_2020} compares the topic proportions of Black females aged 35-50 from the state of Georgia (Group 1) and White males aged 18-34 from the state of California (Group 2). We observe that Topic 2 is more prominent in Group 2 compared to Group 1, while Topic 6 is more prominent in Group 1 compared to Group 2.

\begin{figure}
\begin{center}
\includegraphics[scale=0.6]{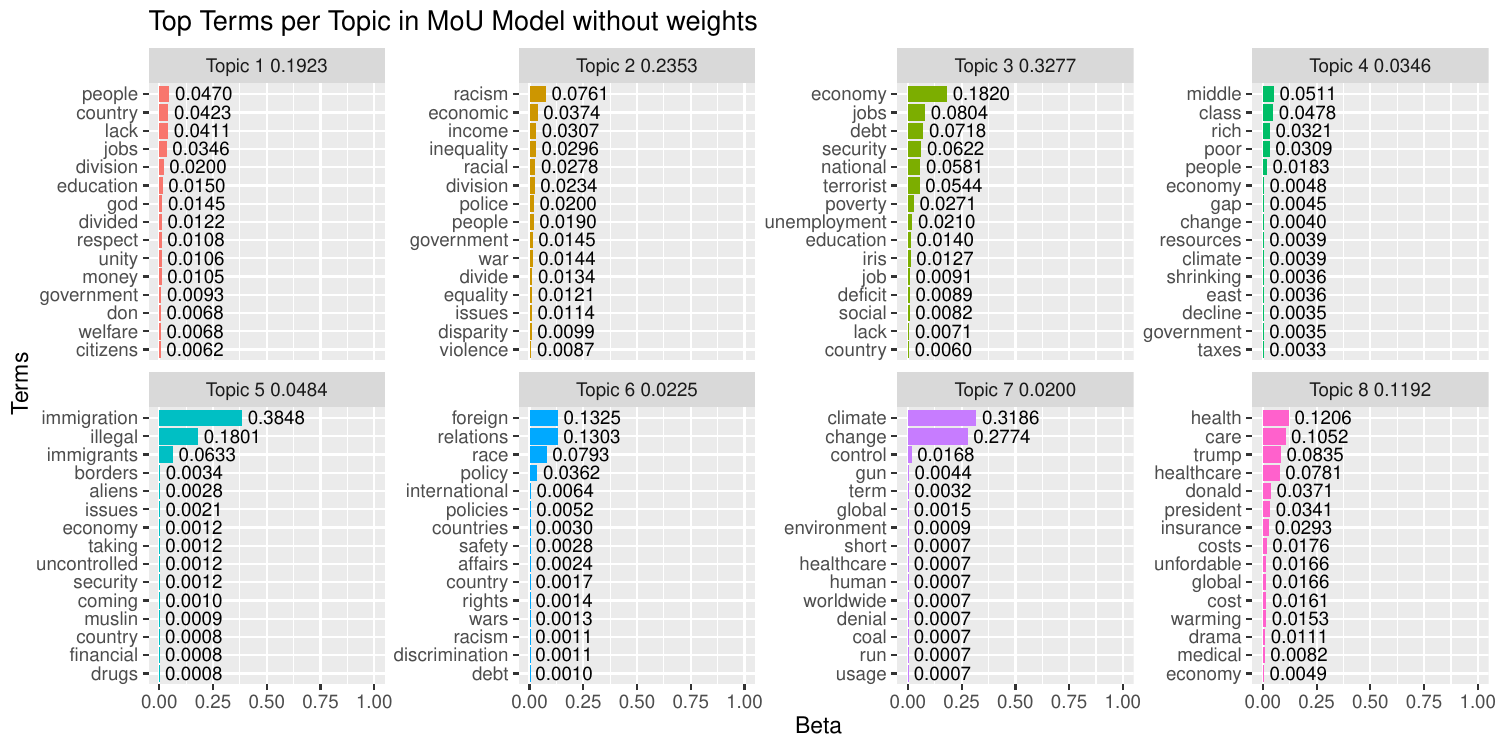}
\caption{Estimated distribution of the top 15 words over each of the $J=8$ latent topics obtained from the MoU model under the non-informative survey design, for the ANES 2016 data.}
\label{fig:beta_plot_2016_1}
\end{center}
\end{figure}

\subsection{Results from the ANES 2016 data}
\begin{figure}
\begin{center}
\includegraphics[scale=0.6]{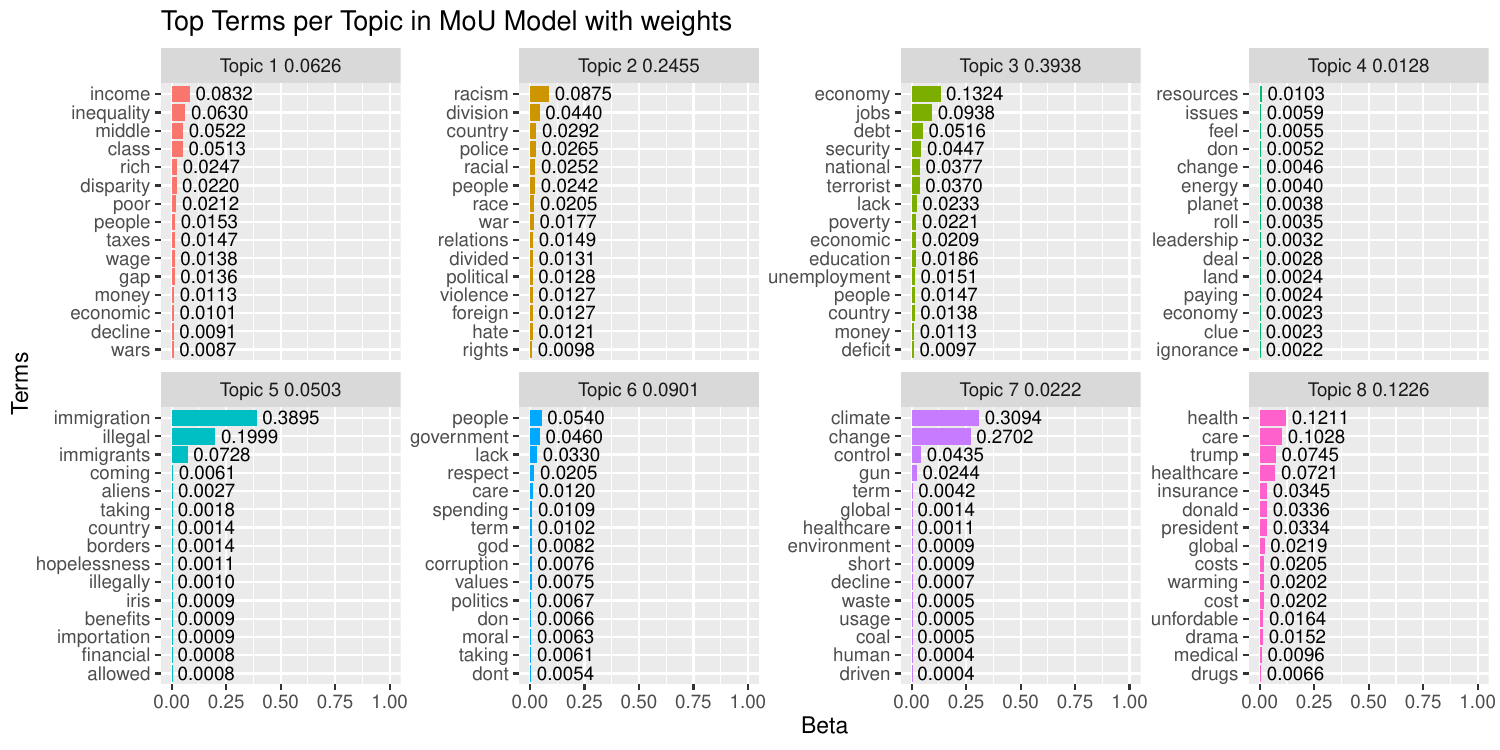}
\caption{Estimated distribution of the top 15 words over each of the $J=8$ latent topics obtained from the MoU model accounting for informative sampling, for the ANES 2016 data.}
\label{fig:beta_plot_2016_2}
\end{center}
\end{figure}

We analyze the the free response texts  from the ANES 2016 data which contains $M= 3215$ responses (or documents) and has vocabulary size  of $V= 1452$. Based on our criterion, the number of topics are chosen to be $J=8$. We run the two models and compare the estimated topic proportions, along with the distribution of the top $15$ words across each topics shown in Figures \ref{fig:beta_plot_2016_1} and \ref{fig:beta_plot_2016_2}. The three most frequent topics based on the distribution of words correspond to $1)$ economy and national security, $2)$ racial and political divisions and, $3)$ healthcare, global warming and political influence as the most important problems in the country. The MoU model accounting for informative sampling estimates the top three topic proportions as $0.3938$, $0.2455$, and $0.1226$ which are underestimated in the standard MoU model as $0.3277$, $0.2353$, and $0.1192$ respectively.

\begin{figure}
\begin{center}
\includegraphics[scale=0.55]{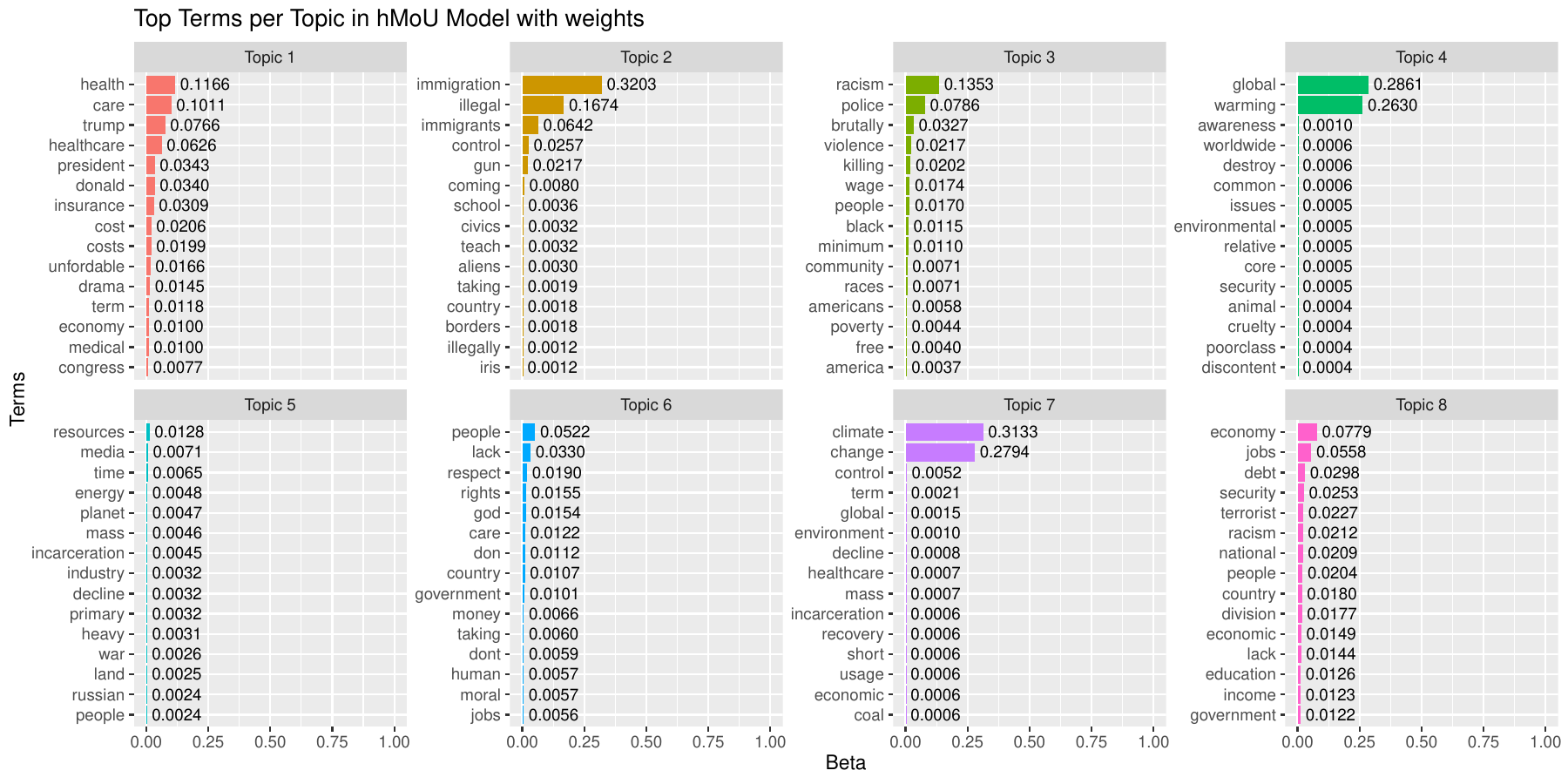}
\caption{Estimated distribution of the top 15 words over each of the $J=8$ latent topics obtained from the hierarchical MoU model under the informative survey design, for the ANES 2016 data.}
\label{fig:hMoU_beta_plot_2016}
\end{center}
\end{figure}

\begin{figure}
\begin{center}
\includegraphics[scale=0.55]{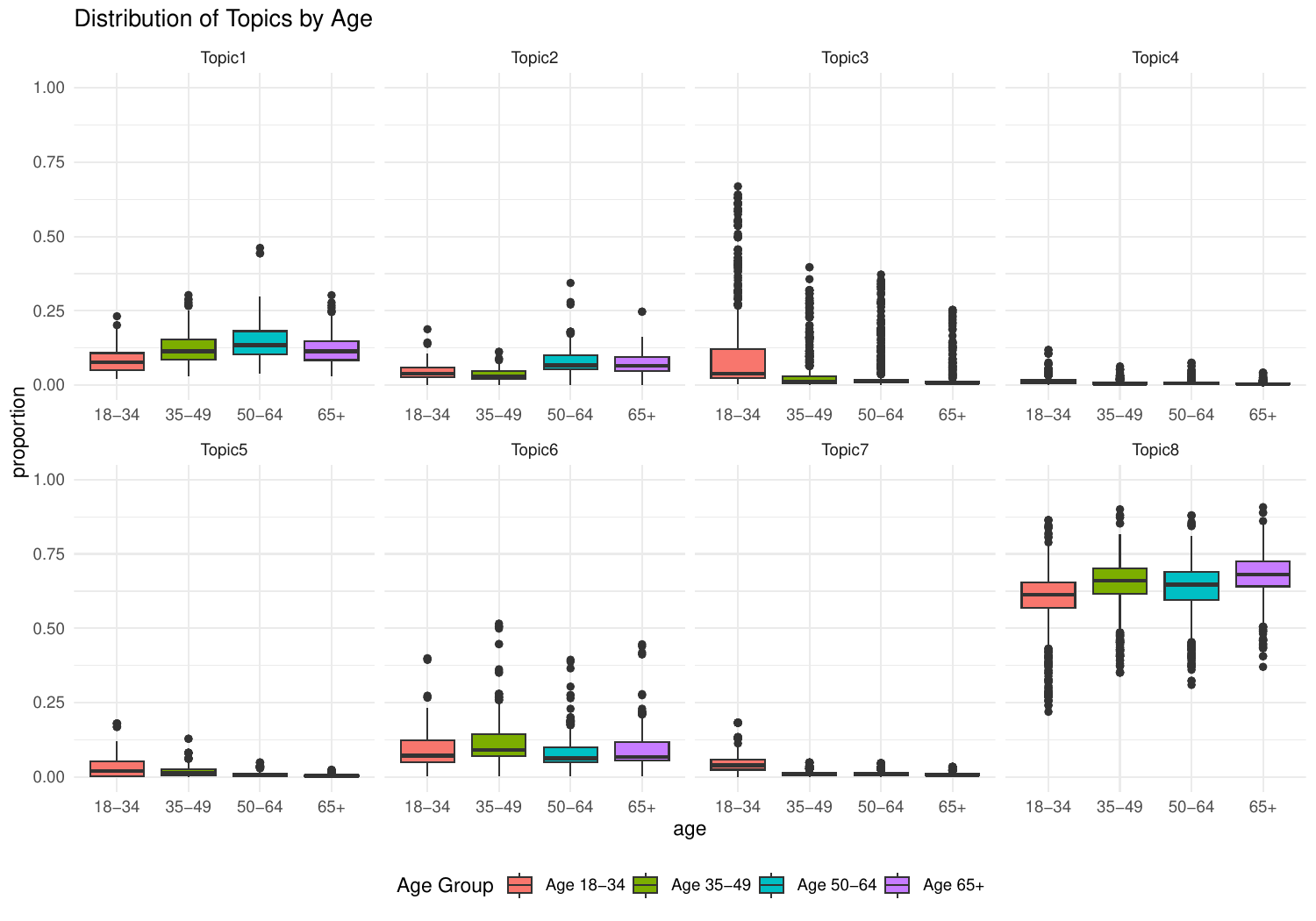}
        \caption{Distribution of topic proportions for each of the $J=8$ latent topics obtained from the hMoU model across different age groups for the ANES 2016 data.}
        \label{fig:topic_by_age_2016}
\end{center}
\end{figure}
\begin{figure}
\includegraphics[scale=0.55]{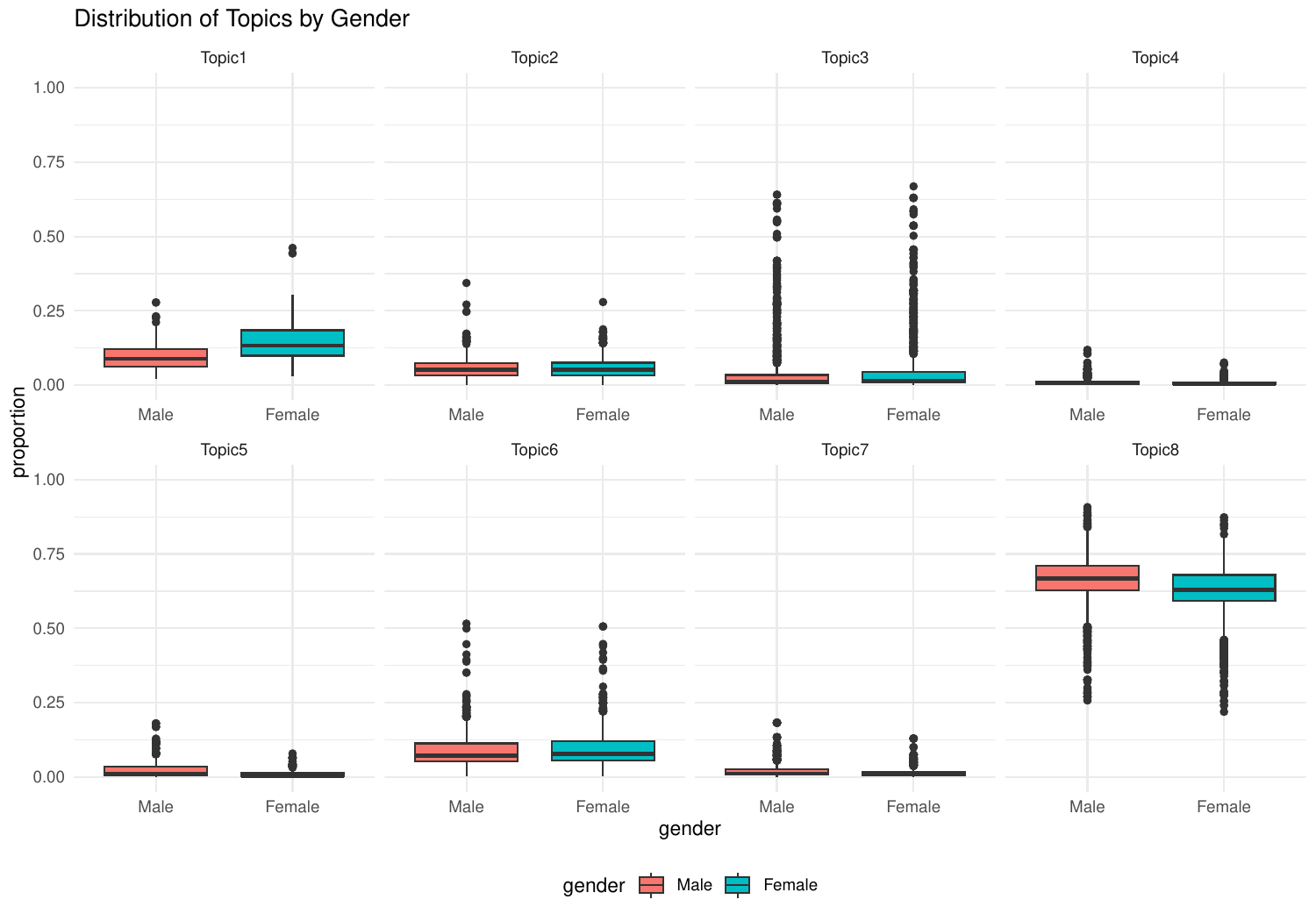}
        \caption{Distribution of topic proportions for each of the $J=8$ latent topics obtained from the hMoU model across different gender groups for the ANES 2016 data.}
        \label{fig:topic_by_gender_2016}
\end{figure}
\begin{figure}
\includegraphics[scale=0.55]{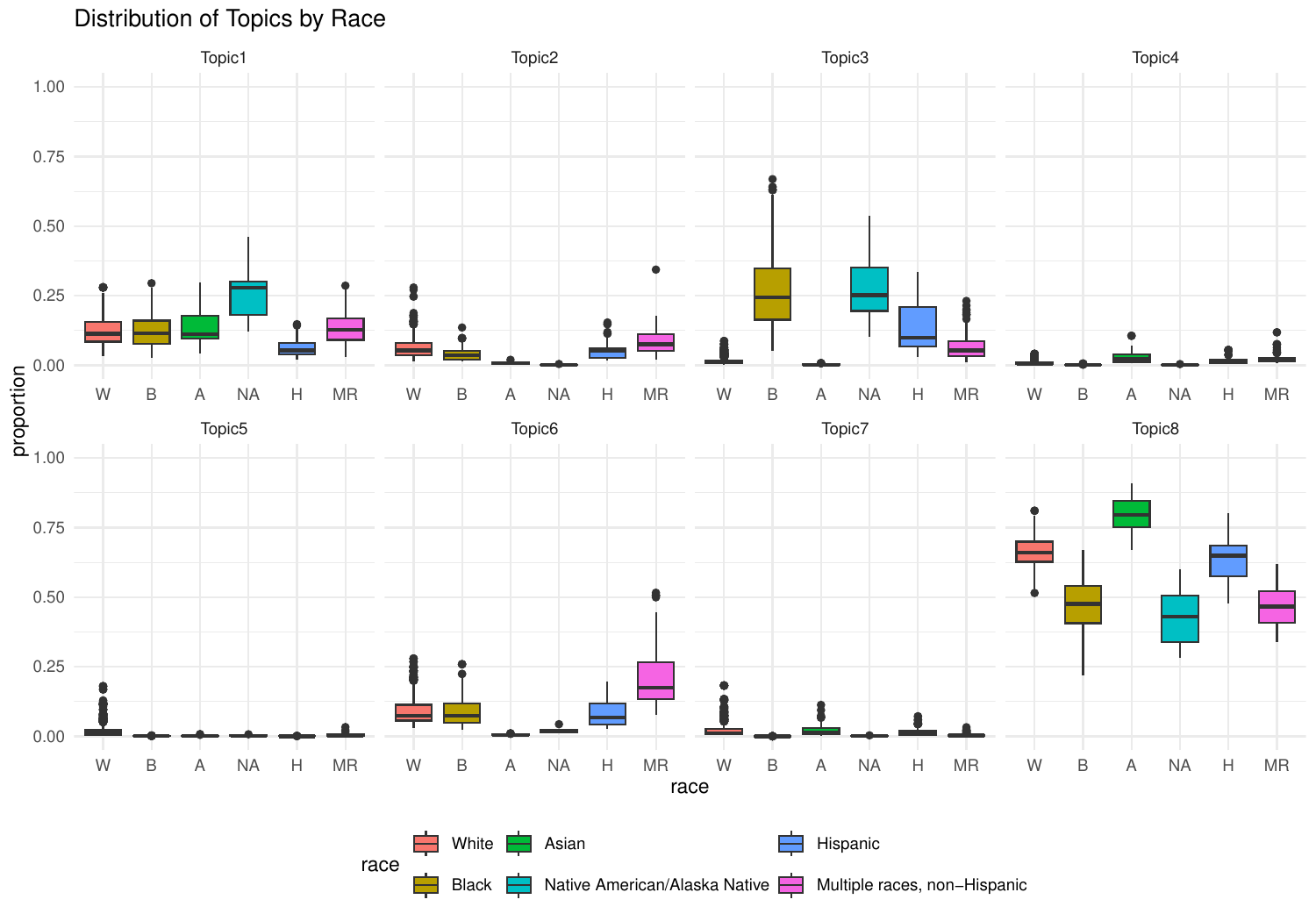}
        \caption{Distribution of topic proportions for each of the $J=8$ latent topics obtained from the hMoU model across different race groups for the ANES 2016 data.}
        \label{fig:topic_by_race_2016}
\end{figure}
\begin{figure}[H]
\begin{center}
\includegraphics[scale=0.55]{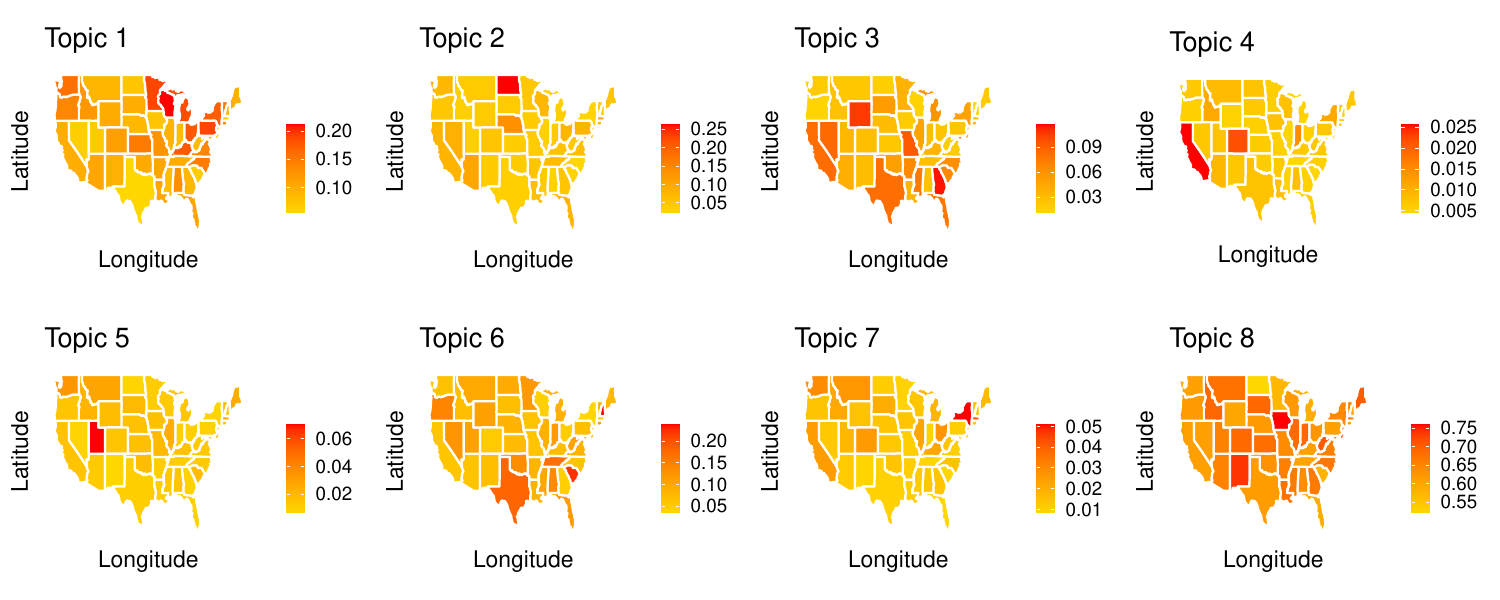}
        \caption{Distribution of topic proportions for each of the $J=8$ latent topics obtained from the hMoU model across different states for the ANES 2016 data.}
        \label{fig:topic_by_state_2016}
\end{center}
\end{figure}

Next, we present the results from the   hierarchical MoU model. Figure~\ref{fig:hMoU_beta_plot_2016} shows the distribution of the top $15$ words across each topic and 
Figures~\ref{fig:topic_by_age_2016}-\ref{fig:topic_by_state_2016} shows the distribution of the $J=8$ topic proportions across various groups including age, gender, race, and state. Similar to the MoU under the informative sampling model (Figure~\ref{fig:beta_plot_2016_2}) the most  frequent topic  (Topic 6) corresponds to \textit{economy and national security} (Figure~\ref{fig:hMoU_beta_plot_2016}). Among the age groups, Topic 3 is more prominent among respondents aged 18-34 compared to other age groups. Across the gender, Topic 1 is more prominent among females than males, while Topic 8 is more prominent among males than females. Across race groups, Topic 1 is most prominent among Native American/Alaska Native respondents; Topic 3 is more prominent among the Black race group; Topic 6 is most prominent among individuals identifying as multiple races, non-Hispanic; and Topic 8 is more prominent among the Asian race group compared to other race groups. Across states, we observe the following: Topic 1 is most prominent among respondents from Wisconsin and Minnesota; Topic 2 is more prominent among respondents from North Dakota; Topic 3 is most prominent among respondents from Wyoming; Topic 4 is most prominent among respondents from California; Topic 5 is most prominent among respondents from Utah; Topic 6 is most prominent among respondents from South Carolina; Topic 7 is most prominent among respondents from New York; and Topic 8 is most prominent among respondents from Iowa and New Mexico; compared to respondents from other states.

\begin{figure}[H]
\begin{center}
\includegraphics[scale=0.5]{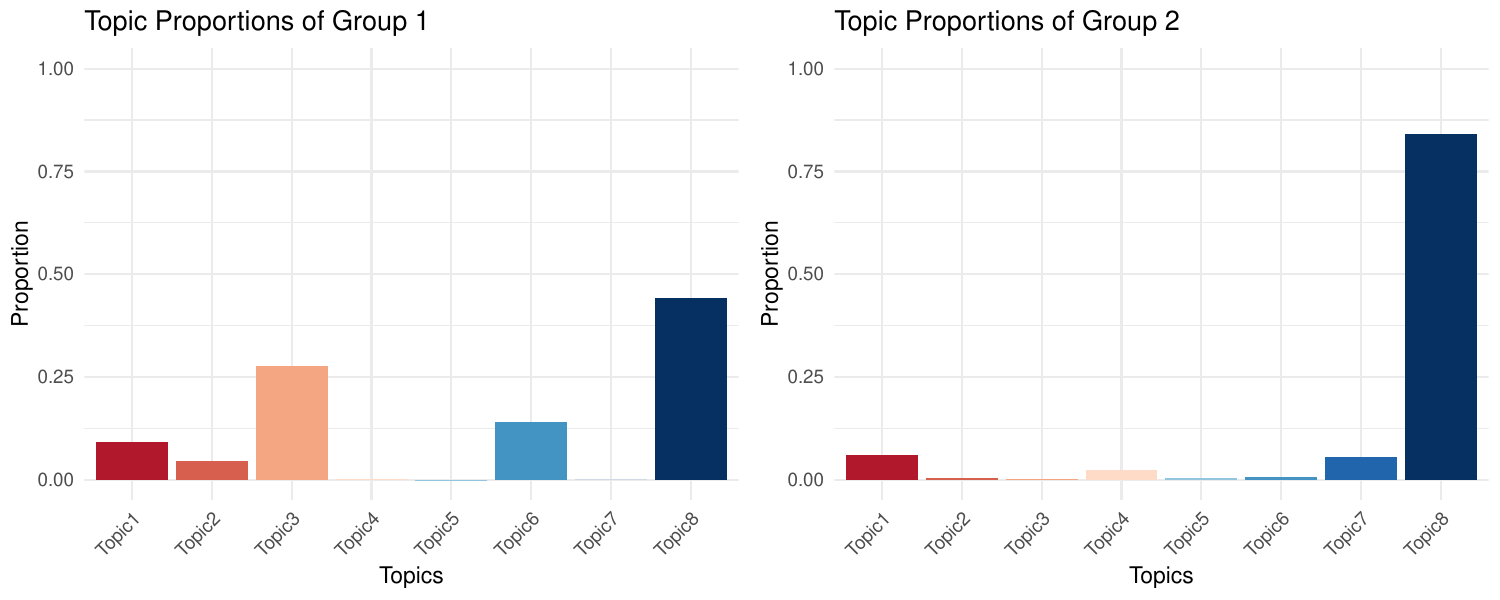}
\caption{Barplots comparing topic proportions of two distince respondent groups. Group 1 consists of 
respondents who are Black females aged 51-64 from the state of Texas and Group 2 consists of respondents who are Asian males aged 18-34 from the state of Illinois.}
\label{fig:compare_2016}
\end{center}
\end{figure}

Further the hierarchical MoU model can also be used to  compare the topic proportions for two distinct respondent groups. For instance, Figure~\ref{fig:compare_2016} compares the topic proportions of 
Black females aged 51-64 from the state of Texas (Group 1) and Asian males aged 18-34 from the state of Illinois (Group 2). We observe that Topics 1 and 6 is more prominent in Group 1 compared to Group 2, while Topic 1 is more prominent in Group 2 compared to Group 1.

\section{Discussion}\label{disc}

In this paper, we extend the standard topic model (Mixture of Unigrams) to handle  informative sampling. This can be particularly useful in the context of survey data where data is usually collected on individual characteristics. Our method incorporates a survey-weighted pseudo-likelihood in order to account for the sample selection probabilities to estimate the parameters of interest, thus avoiding biased estimates. We illustrate the effectiveness of our approach in comparison to the standard model using a simulation study where the informative sample from the population is drawn such that there exists an underlying relationship between sample selection probabilities and the latent topic structure. From the simulation study, we observe that the MoU model accounting for sampling design leads to more reliable parameter estimation as opposed to the standard model, which results in biased estimates. We also introduce a  hierarchical MoU model, where the the topic proportions are modeled as functions of document-level fixed and random effects. We demonstrate the performance of the MoU and the hMoU models accounting for informative sampling on two datasets containing post-election responses from the ANES that are publicly available and  extract meaningful topics based on the responses on the most important problems in the country. The hierarchical MoU model, allows us to compare topic proportions  across fixed effects, including gender, race, and age groups, along with state-level random effects. Both our models  can be used to any complex survey datasets with open ended responses.

\section*{Acknowledgments}
  This article is released to inform interested parties of ongoing research and to encourage discussion. The views expressed on statistical issues are those of the authors and not those of the NSF or U.S. Census Bureau. This research was partially supported by the U.S. National Science Foundation (NSF) under NSF grants NCSE-2215168 and NCSE-2215169.

\bibliographystyle{apalike}
\bibliography{example}

\end{document}